\documentclass[iop]{emulateapj}

\usepackage{pbox}
\usepackage{color}
\usepackage{textcomp}

\shorttitle{Physical properties of MACS0647-JD}
\shortauthors{Lam et al.}

\begin{document}

\title{Detection of a Lensed z$\approx$11 Galaxy in the Rest-Optical with Spitzer/IRAC\\and the Inferred SFR, Stellar Mass, and Physical Size}

\author{Daniel Lam\altaffilmark{1}, 
        Rychard J. Bouwens\altaffilmark{1}, 
        Dan Coe\altaffilmark{2}, 
        Adi Zitrin\altaffilmark{3}, 
        Christopher Barber\altaffilmark{1}, 
        Ivo Labb\'{e}\altaffilmark{4},
        Themiya Nanayakkara\altaffilmark{1},
        Megan Donahue\altaffilmark{5},
        Renske Smit\altaffilmark{6,7},
        Xinwen Shu\altaffilmark{8},
        Ranga-Ram Chary\altaffilmark{9},
        John Moustakas\altaffilmark{10},
        Mario Nonino\altaffilmark{11}, 
        Daniel D. Kelson\altaffilmark{12},
        Tom Broadhurst\altaffilmark{13,14,15},
        Larry Bradley\altaffilmark{2},
        Mauricio Carrasco\altaffilmark{16},
        Piero Rosati\altaffilmark{17}}
\affil{$^{1}$ Leiden Observatory, Leiden University, NL-2300 RA Leiden, The Netherlands\\
       $^{2}$ Space Telescope Science Institute, Baltimore, MD, USA\\
       $^{3}$ Physics Department, Ben-Gurion University of the Negev, P.O. Box 653, Beer-Sheva 8410501, Israel\\
       $^{4}$ Centre for Astrophysics and SuperComputing, Swinburne, University of Technology, Hawthorn, Victoria, 3122, Australia\\
       $^{5}$ Physics \& Astronomy Department, Michigan State University, East Lansing, MI 48824-2320, USA\\
       $^{6}$ Cavendish Laboratory, University of Cambridge, 19 JJ Thomson Avenue, Cambridge CB3 0HE, UK\\
       $^{7}$ Kavli Institute for Cosmology, University of Cambridge, Madingley Road, Cambridge CB3 0HA\\
       $^{8}$ Department of Physics, Anhui Normal University, Wuhu, Anhui, 241000, Peopleʼs Republic of China\\
       $^{9}$ MS314-6, Infrared Processing and Analysis Center, California Institute of Technology, Pasadena, CA 91125, USA\\
       $^{10}$ Department of Physics and Astronomy, Siena College, 515 Loudon Road, Loudonville, NY 12211\\
       $^{11}$ INAF – Osservatorio Astronomico di Trieste, via G. B. Tiepolo 11, I-34143, Trieste, Italy\\
       $^{12}$ The Observatories of the Carnegie Institution for Science, 813 Santa Barbara Street, Pasadena, CA 91101, USA\\
       $^{13}$ Department of Theoretical Physics, University of The Basque Country UPV/EHU, E-48080 Bilbao, Spain\\
       $^{14}$ Donostia International Physics Center (DIPC), 20018 Donostia, The Basque Country\\
       $^{15}$ IKERBASQUE, Basque Foundation for Science, E-48013 Bilbao, Spain\\
       $^{16}$ Zentrum fur Astronomie, Institut f\"{u}r Theoretische Astrophysik, Philosophenweg 12, 69120, Heidelberg, Germany\\
       $^{17}$ Dipartimento di Fisica e Scienze della Terra, Universit\'{a} degli Studi di Ferrara, Via Saragat 1, I-44122 Ferrara, Italy
       }

\begin{abstract}
We take advantage of new 100-hour Spitzer/IRAC observations available
for MACS0647-JD, a strongly lensed $z\approx11$ galaxy candidate, to provide improved
constraints on its physical properties.  Probing the physical
properties of galaxies at $z>8$ is challenging due to the inherent
faintness of such sources and the limited wavelength coverage
available.  Thanks to the high
$\approx$2-6$\times$ lensing magnification of the multiple images of
MACS0647-JD, we can use the sensitive Spitzer/IRAC data to probe the
rest-frame optical fluxes of MACS0647-JD and investigate its physical
properties including the age and the stellar mass.  In deriving Spitzer/IRAC fluxes for
MACS0647-JD, great care is taken in coping with the impact of three
bright ($\approx$8-16 mag) stars in our field to ensure robust results.
Assuming a constant star formation rate, the age, stellar mass, and rest-frame UV slope we estimate for
MACS0647-JD based on a stack of the photometry are log$_{10}$(age/yr) = 8.6$^{+0.1}_{-2.1}$, log$_{10}$(M$_{*}$/M$_{\odot}$) = 9.1$^{+0.2}_{-1.4}$, and $\beta = -$1.3$\pm$0.6, 
respectively.  We compare our results with expectations from the EAGLE simulation and find that MACS0647-JD has properties consistent with corresponding to the most massive and rapidly star-forming galaxies in the simulation.  We also find that its radius, 105$\pm$28 pc, is a factor of $\approx$2 smaller than the mean size in a separate simulation project DRAGONS.  Interestingly enough, the observed size is similar to the small sizes seen in very low-luminosity $z\approx6$-10 galaxies behind lensing clusters.
\end{abstract}


\section{Introduction}

One of the most important frontiers in observational astronomy
involves the identification and characterization of galaxies at early
epochs in the universe.  Star-forming galaxies are suspected to have
an important role in the reionization of the universe
\citep{stanway03,yan04,oesch09,ouchi10,bouwens12,bouwens15,finkelstein12,robertson13,robertson15,ishigaki15,mitra15}.
While hundreds of galaxies are known at z$\approx$6-8
\citep{bouwens11,bouwens15,mclure13,schenker13,schmidt14,bradley14,finkelstein15,stark16, livermore17, ishigaki18},
samples of $z\approx9$ and $z\approx10$ galaxies are much more modest, with
only $\approx$30 candidates known to the present
\citep{bouwens11nat,bouwens14,bouwens15,bouwens16,zheng12,oesch13,coe13,zitrin14,ishigaki15,mcleod15,mcleod16,laporte16}.
This is particularly the case at $z>10$ \citep{coe13,ellis13,oesch16}.

Despite the successful identification of modest samples of galaxy candidates in
the $z\approx9$-11 universe using the Lyman Break, detailed
characterization of these sources is much more difficult.  This is as
a result of the stellar continuum being largely redshifted out of
wavelength range to which the Hubble Space Telescope (HST) is sensitive.
Observations at redder wavelengths, such as those provided by the
Spitzer Space Telescope, provide a much better probe of the stellar
continuum.  An example of such observations is the Spitzer Ultra Faint
Survey Program \citep[SURFS'UP:][]{bradac14}, which targets high
redshift galaxies lensed by 10 galaxy clusters.  Using the SURFS'UP
data, \cite{huang16} not only report the successful detection of thirteen 6
$<$ z $<$ 10 galaxies, but then proceed to derive physical
characteristics for the same sources.

It is interesting to make use of Spitzer/IRAC observations to improve
our characterization of sources not just at $z<8$ but also at $z>8$.  One
such demonstration was provided by \cite{zheng12} showing the
existence of a possible Balmer break in the Spitzer/IRAC photometry
\citep[see also][]{bradac14, hoag18, hashimoto18} while the analysis by \cite{oesch16} of
the $z=11.1$ galaxy, GN-z11, showed a relatively blue stellar continuum with no
evidence for a break.  \cite{oesch16} estimate a stellar mass of
$10^{9.0\pm0.4}$ $M_{\odot}$ for GN-z11 and a $UV$-continuum slope of
$-2.5\pm0.2$.  \cite{wilkins16} make use of the available
HST+Spitzer/IRAC observations for five bright $z\approx9$-10 galaxy candidates to
estimate the mean $UV$-continuum slope at $z\approx10$, finding a value
of $-2.1\pm0.2\pm0.3$.

One other particularly interesting $z>10$ galaxy candidate where one could
pursue an improved characterization of the source is MACS0647-JD.
Discovered by \cite{coe13}, it is a z$\approx$11 galaxy candidate triply lensed by
the CLASH cluster MACS0647.  The three multiple images of MACS0647-JD,
i.e., JD1, JD2, and JD3, are estimated to be magnified by
$\approx$6.0$\times$, $\approx$5.5$\times$, and $\approx$2.1$\times$,
respectively.  MACS0647-JD is only detected in the two HST WFC3/IR
filters at the longest wavelengths, namely F140W and F160W.  Only 
one of the three multiple images was detected in the Spitzer/IRAC 4.5$\mu$m
data (5-hour depth each in 3.6 $\mu$m and 4.5 $\mu$m band) of the Spitzer Lensing
Survey (PI: Egami).  
The low IRAC fluxes strongly disfavored the possibility of
MACS0647-JD being a z$\approx$2.5 red galaxy.   Furthermore, as shown by \cite{pirzkal15}, the lack of detectable emission lines in the HST grism data (which are expected if MACS0647-JD is actually a lower-redshift interloper) strengthens the case for MACS0647-JD corresponding to a z$\approx$11 galaxy.

In order to better constrain the nature of MACS0647-JD as well as its
physical properties, we obtained much deeper observations on MACS0647
with Spitzer/IRAC, increasing the exposure time by a factor of 10 to
50 hours per band.  Such long integration would allow us to securely
detect the source, measure its UV-continuum slope, and also provide a
glimpse into other physical properties of such sources like their
stellar mass.  This is interesting since MACS0647-JD and GN-z11
\citep{bouwens10,oesch16} are the only examples of sources we know
which existed just 400 Myr after the Big Bang in the heart of the
reionization epoch.

The plan for the paper is as follows.  \S\ref{sec:observation}
provides a description of the observational data for MACS0647-JD and
the steps taken to cope with the presence of bright stars around
MACS0647, including the subtraction of the stars and handling of
artifacts resulting from the bright stars.  \S\ref{sec:photometry}
describes our procedures for doing photometry in crowded fields and 
making size measurements.
\S\ref{sec:sps} discusses the stellar population properties for
MACS0647-JD we infer, and \S\ref{sec:eagle} compare our findings with
the predictions of a cosmological simulation.  Throughout the paper,
we adopt the traditional concordance cosmology with $\Omega_{M}$ =
0.3, $\Omega_{\Lambda}$ = 0.7, and $H_{0}$ = 70 km s$^{-1}$.  All
magnitudes are in the AB system \citep{oke83}.

\begin{deluxetable*}{cccc}
\tablecaption{\label{table:observation} Wavelength coverage, exposure
  times, and sensitivities of the available HST+Spitzer/IRAC
  observations available in various bandpasses over MACS0647.  Adopted and modified from
  \cite{coe13}. } \tablehead{ \colhead{Filter} & \colhead{wavelength
    coverage ($\micron$)} & \colhead{exposure time (s)} &
  \colhead{3$\sigma$ limiting AB magnitude$^{1}$} } \startdata 
  F225W & 0.20 - 0.30 & 3805 & 26.8 \\
F275W & 0.23 - 0.31 & 3879 & 26.8 \\ 
F336W & 0.30 - 0.37 & 2498 & 26.7 \\ 
F390W & 0.33 - 0.45 & 2545 & 26.6\\ 
F435W & 0.36 - 0.49 & 2124 & 27.2 \\ 
F475W & 0.39 - 0.56 & 2248 & 27.7 \\ 
F555W & 0.46 - 0.62 & 7740 & 28.4 \\ 
F606W & 0.46 - 0.72 & 2064 & 27.1 \\ 
F625W & 0.54 - 0.71 & 2131 & 26.7 \\ 
F775W & 0.68 - 0.86 & 2162 & 26.5 \\ 
F814W & 0.69 - 0.96 & 12760 & 28.2 \\ 
F850LP & 0.80 - 1.09 & 4325 & 26.5 \\ 
F105W & 0.89 - 1.21 & 2914 & 28.0 \\ 
F110W$^{2}$ & 0.88 - 1.41 & 1606 & 28.0 \\ 
F125W & 1.08 - 1.41 & 2614 & 27.8 \\ 
F140W & 1.19 - 1.61 & 2411 & 27.9 \\ 
F160W & 1.39 - 1.70 & 5229 & 27.9 \\ 
IRAC [3.6]$^{3}$ & 3.10 - 4.00 & 90914 & 25.0 \\
IRAC [4.5]$^{3}$ & 3.90 - 5.10 & 92740 & 25.5 
\enddata
\tablenotetext{1}{Measured from the standard deviation of fluxes in 1000 randomly placed, non-overlapping circular apertures on the background. The aperture radii are 0.2" and 0.9" for HST and Spitzer/IRAC data, respectively. }
\tablenotetext{2}{The exposure affected by earthshine is excluded. }
\tablenotetext{3}{$\approx$50\% of the exposure time is not included in the Spitzer/IRAC images we use for the photometry of each candidate due to ``column-pull down'' artifacts that impact the flux measurements at the position of the various multiple images (Figure~\ref{fig:stars_subtracted_jd1_jd2} and $\S$2.3).}
\end{deluxetable*}

\section{Observational data}\label{sec:observation}

\subsection{Basic Description}

The HST data we utilize for this analysis is identical to those used
in \cite{coe13}.  The publicly available HST imaging taken by the CLASH program
for MACS0647 provides coverage in 17 filters from the UV at $\approx$200
nm to the near-infrared at $\approx$1.6 $\micron$.  Table
\ref{table:observation} indicates the wavelength coverage, exposure
times, and 3-$\sigma$ sensitivities provided by each filter.
Consistent with the treatment in \cite{coe13}, we do not include the
second epoch of observations in the F110W band because of the
significantly elevated noise in those observations which results from
Helium emission in the Earth\textquotesingle s atmosphere \citep{brammer14}. 

For the Spitzer/IRAC data, we make use of the 50-hour exposures we
obtained as a result of an approved program in cycle 7 (PI: Coe) to
obtain improved constraints on the physical properties of MACS0647-JD.
The data are reduced by the same custom procedure described in
\cite{labbe15}, which removes background structures, corrects for
artifacts, masks persistence, rejects cosmic rays, calibrates the
astrometry, and creates mosaic science images.  Details of the data
reduction process can be found in \cite{labbe15}.

The new Spitzer/IRAC observations are combined with the previous
observation (5 hours per band, PI: Egami) that was only available then
to \cite{coe13}, bringing the total exposure time to 55 hours per
band.

\begin{figure*}
\centering
\includegraphics[width=160mm]{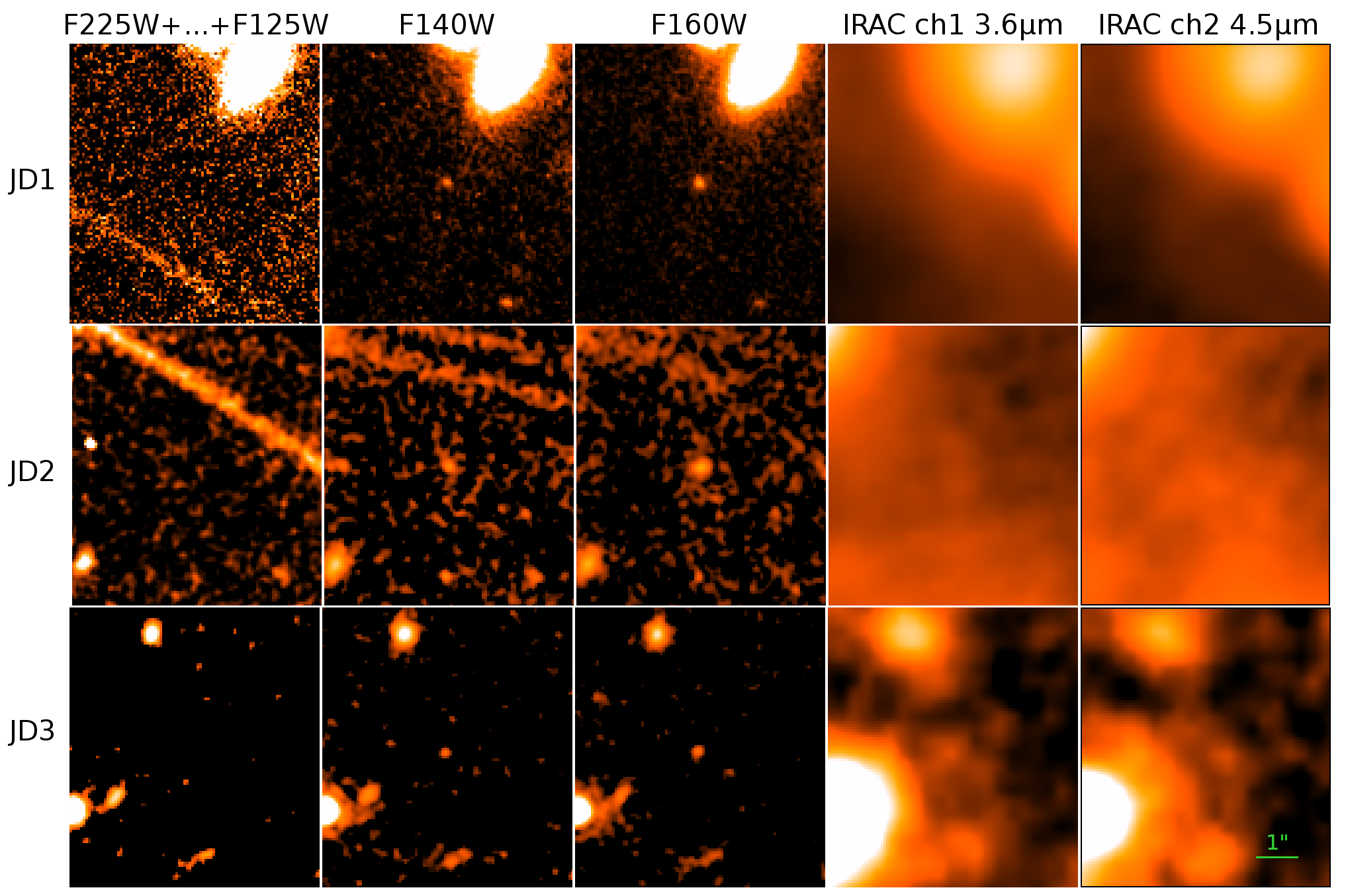}
\caption{ Image stamps (3.5$''$ $\times$ 3.5$''$) of the three
  different multiple images of the $z\approx11$ candidate MACS0647-JD
  presented in the upper (JD1), middle (JD2), and lower rows (JD3).
  The five columns show left to right (1) the weighted average of the
  available HST data from the F225W band to the F125W band (15 bands),
  (2) the HST WFC3/F140W image, (3) the HST WFC3/F160W image, (4) the
  Spitzer IRAC 3.6 $\mu$m image, and the Spitzer IRAC 4.5 $\mu$m
  image.  The panels are not displayed in the same brightness scale
  but are adjusted for better viewability.  The very bright nearby stars are not subtracted (as discussed in Section \ref{sec:subtract_stars}) in this figure. }
\label{fig:postage_stamps}
\end{figure*}

The Spitzer/IRAC data we utilize represents a ten-fold increase in the
exposure time over that which was available to \cite{coe13}.  The
sensitivity of the Spitzer/IRAC observations are determined based on
the noise fluctuations in circular apertures of radius $\approx$3.3" over
a blank region away from the cluster center.  
Figure \ref{fig:postage_stamps} shows the image stamps of the three multiple images in both HST and Spitzer/IRAC data. 

\subsection{Subtraction of the Bright Stars Around MACS0647}
\label{sec:subtract_stars}

One particularly challenging aspect of doing photometry on the
$z\approx11$ galaxy behind MACS0647 is the presence of an 8th magnitude
star only 2 arcminutes away from the cluster core, and 13th, 14th,
and 16th magnitude stars in the same $2'$ $\times$ $2'$ WFC3/IR
field.

Clearly, the first step in measuring the Spitzer/IRAC fluxes in the
faint $z\approx11$ multiple images around MACS0647 is the removal of the
extended light of nearby Galactic bright stars.  Otherwise, our
photometric modeling package \textsc{mophongo} will attempt to model
the extended stellar light from the bright stars as originating from
various bright sources nearby (within a radius of 12") the $z\approx11$ multiple images.

\begin{figure}
\centering
\includegraphics[width=85mm]{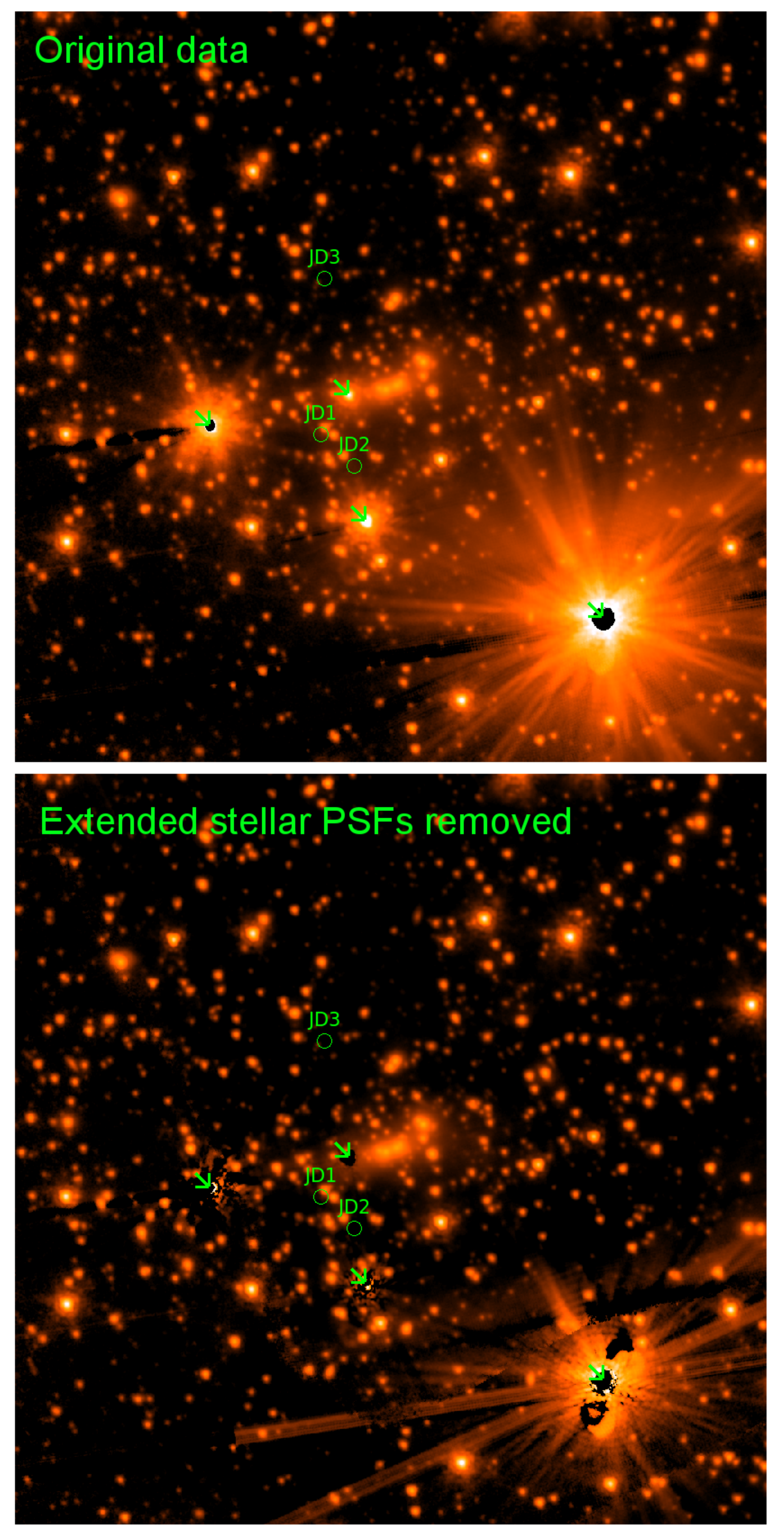}
\caption{(\textit{upper}) The original Spitzer IRAC 3.6$\mu$m image
  with the four bright stars to be removed, with approximate apparent
  magnitudes of 8, 13, 14, and 16 (\textit{indicated by the green
    arrows}).  (\textit{lower}) The same image as the top panel, but
  with the extended stellar light subtracted.  Bright foreground
  sources (mostly cluster galaxies) and saturated stellar cores are
  masked out to allow for an accurate fit to the extended component of
  stellar light from the bright stars.\label{fig:stars_subtracted}}
\end{figure}

Figure \ref{fig:stars_subtracted} shows the four stars that contribute
most of the contamination.  A high dynamical range, extended PSF
model produced by IRSA is combined from observations of stars across a wide range of
fluxes.  The FITS files can be obtained at this
link.\footnote{http://irsa.ipac.caltech.edu/data/SPITZER/docs/irac/\\calibrationfiles/psfprf/}
We then simply use \textsc{galfit} \citep{peng10} to model and
subtract the extended stellar light while masking out the saturated
stellar cores as well as all bright objects in the field of view.  This process is repeated for exposures lying at a range of roll angles.

\subsection{``Column Pull-Down'' Artifact}

A second artifact that impacts the fidelity of the Spitzer/IRAC
imaging observations is the column pull-down artifact,\footnote{Refer
  to the Spitzer IRAC instrument handbook} which is relatively large
in the case of the MACS0647 cluster data due to the brightness of the
foreground stars.  These artifacts reduce the intensities in entire
columns where the bright objects are located.  The Spitzer IRAC basic
calibrated data (BCD) pipeline partially corrects for the column
pull-downs.  However, correcting for this artifact perfectly is
extremely challenging, particularly when working with fields with
extremely bright sources, such as is the case near the centers of
galaxy clusters or where a field has many bright stars such as with
MACS0647.

\begin{figure}
\centering
\includegraphics[width=70mm]{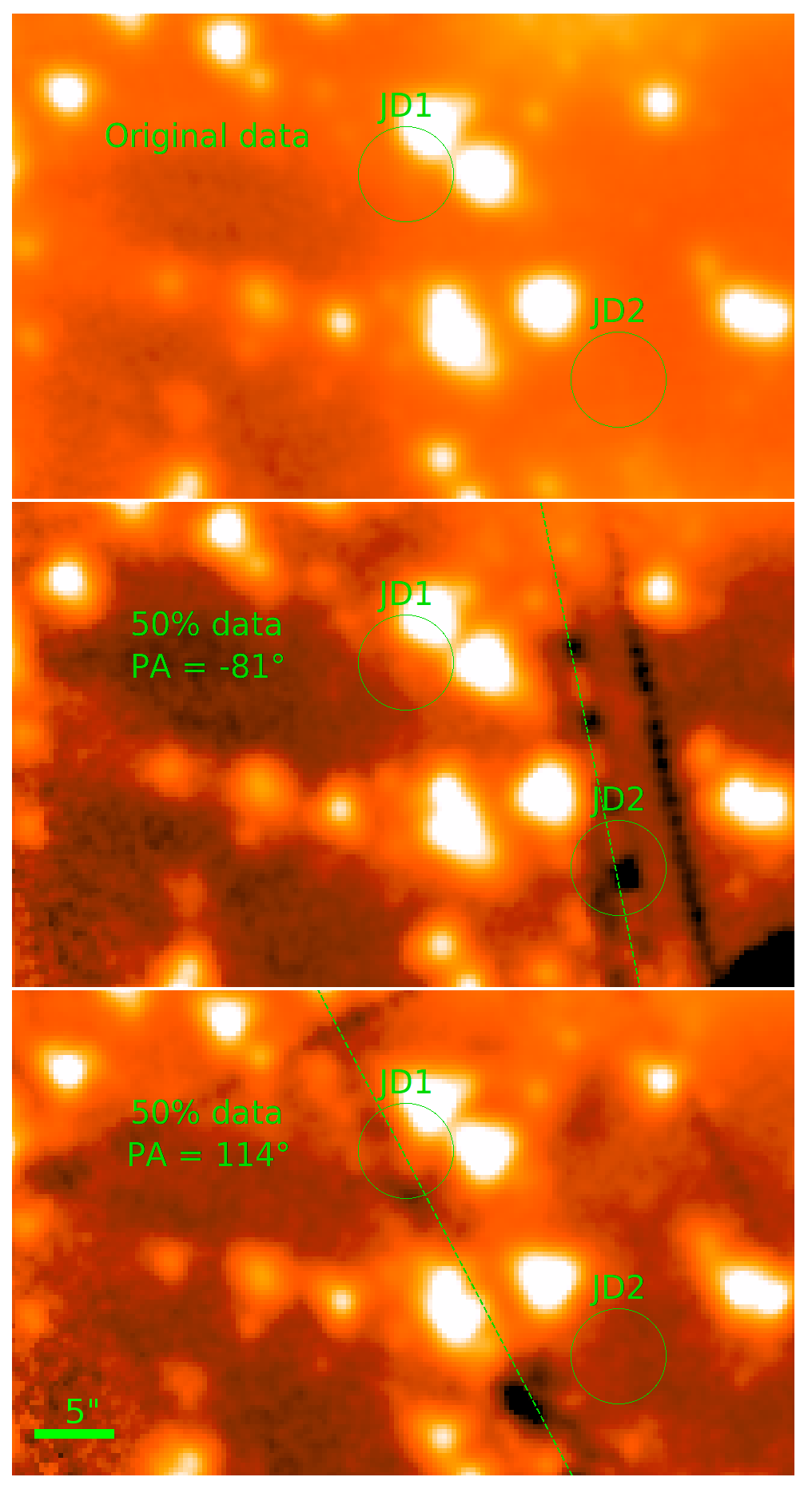}
\caption{ Illustration of the Spitzer/IRAC observations available over
  various multiple images of MACS0647-JD showing the impact that
  ``column pull-down'' artifacts from bright stars have on the images
  (even after correction).  This artifact is especially apparent after
  subtraction of the extended stellar light from original data
  (\textit{upper panel}).  Half of the Spitzer/IRAC exposures are such
  that the ``column pull-down'' artifact runs through JD2
  (\textit{middle panel}) and JD3, while the other half of the
  exposures are such that this same artifact runs through JD1
  (\textit{lower panel}).  We therefore only make use of that half of
  the Spitzer/IRAC observations unaffected by the artifact in
  measuring fluxes for the individual sources.
\label{fig:stars_subtracted_jd1_jd2}}
\end{figure}

Based on a simple visual inspection of the individual exposures
acquired at different roll angles of the telescope, we found that half
of the data suffered from column pull-down artifacts oriented such
that the first multiple image, JD1, is affected, while for the other
half of the data, the second and third multiple images JD2 and JD3 are
affected.  Figure \ref{fig:stars_subtracted_jd1_jd2} shows how column
pull-down artifact overlaps with the position of JD1 in half of the
data while in the other half of the data JD2 and JD3 are impacted.
This unfortunate coincidence of various multiple images of MACS0647-JD
with ``pull-down'' artifacts reduces the amount of usable data by a
factor of $\approx$2 for each multiple image. We construct one mosaic image for JD1 from exposures with roll angles such that JD1 is not affected by the "column pull-down" artifact, and similarly we construct a second mosaic image for JD2 and JD3 from exposures with roll angles such these two multiple images are not affected by the "column pull-down" artifact.

\begin{deluxetable*}{ccccc}
\tablecaption{\label{table:photometry} Coordinates, estimated magnification factors, and demagnified photometry\tablenotemark{1} of the three multiple images of MACS0647-JD: JD1, JD2, and JD3.}
\tablehead{
           \colhead{ } & 
           \colhead{JD1} & 
           \colhead{JD2} & 
           \colhead{JD3} & 
           \colhead{combined}
           }
\startdata
RA (J2000) & 06:47:55.731 & 06:47:53.112 & 06:47:55.452 & \\
Decl. (J2000) & +70:14:35.76 & +70:14:22.94 & +70:15:38.09 & \\
\pbox{20cm}{Magnification \\ (ZITRIN LTM)} & 6.0$^{+0.6}_{-0.7}$ & 5.5$^{+1.0}_{-0.4}$ & 2.1$^{+0.3}_{-0.1}$ & 13.6$^{+1.2}_{-0.8}$ \\
 F225W  &   $-22 \pm   9$ ($-2.4 \sigma$)  &   $ -7 \pm   9$ ($-0.8 \sigma$)  &   $  6 \pm  15$ ($ 0.4 \sigma$)  &   $ -8 \pm   7$ ($-1.2 \sigma$)  \\
 F275W  &   $-16 \pm   9$ ($-1.8 \sigma$)  &   $ -6 \pm   8$ ($-0.7 \sigma$)  &   $ 23 \pm  12$ ($ 2.0 \sigma$)  &   $  1 \pm   5$ ($ 0.1 \sigma$)  \\
 F336W  &   $  0 \pm   6$ ($ 0.1 \sigma$)  &   $  9 \pm   5$ ($ 1.7 \sigma$)  &   $-12 \pm   9$ ($-1.4 \sigma$)  &   $ -1 \pm   4$ ($-0.2 \sigma$)  \\
 F390W  &   $ -1 \pm   3$ ($-0.4 \sigma$)  &   $  0 \pm   3$ ($ 0.1 \sigma$)  &   $  0 \pm   5$ ($ 0.1 \sigma$)  &   $ -0 \pm   2$ ($-0.1 \sigma$)  \\
 F435W  &   $  0 \pm   4$ ($ 0.0 \sigma$)  &   $  8 \pm   4$ ($ 1.7 \sigma$)  &   $  2 \pm   7$ ($ 0.4 \sigma$)  &   $  3 \pm   3$ ($ 1.1 \sigma$)  \\
 F475W  &   $ -0 \pm   2$ ($-0.1 \sigma$)  &   $ -5 \pm   3$ ($-1.6 \sigma$)  &   $  3 \pm   4$ ($ 0.9 \sigma$)  &   $ -1 \pm   2$ ($-0.4 \sigma$)  \\
 F555W  &   $ -0 \pm   2$ ($-0.3 \sigma$)  &   $  2 \pm   1$ ($ 1.7 \sigma$)  &   $  3 \pm   2$ ($ 1.5 \sigma$)  &   $  2 \pm   1$ ($ 1.6 \sigma$)  \\
 F606W  &   $  0 \pm   3$ ($ 0.2 \sigma$)  &   $  2 \pm   4$ ($ 0.6 \sigma$)  &   $ -0 \pm   3$ ($-0.2 \sigma$)  &   $  1 \pm   2$ ($ 0.4 \sigma$)  \\
 F625W  &   $ -6 \pm   4$ ($-1.6 \sigma$)  &   $ -9 \pm   5$ ($-2.1 \sigma$)  &   $ 11 \pm   5$ ($ 2.3 \sigma$)  &   $ -1 \pm   3$ ($-0.6 \sigma$)  \\
 F775W  &   $  1 \pm   5$ ($ 0.1 \sigma$)  &   $ -3 \pm   9$ ($-0.3 \sigma$)  &   $  2 \pm   5$ ($ 0.4 \sigma$)  &   $ -0 \pm   4$ ($-0.0 \sigma$)  \\
 F814W  &   $  0 \pm   1$ ($ 0.0 \sigma$)  &   $ -0 \pm   1$ ($-0.4 \sigma$)  &   $ -1 \pm   1$ ($-0.7 \sigma$)  &   $ -0 \pm   1$ ($-0.6 \sigma$)  \\
F850LP  &   $ -0 \pm   5$ ($-0.1 \sigma$)  &   $  0 \pm   5$ ($ 0.0 \sigma$)  &   $  3 \pm   7$ ($ 0.4 \sigma$)  &   $  1 \pm   3$ ($ 0.2 \sigma$)  \\
 F105W  &   $  2 \pm   2$ ($ 0.9 \sigma$)  &   $  3 \pm   2$ ($ 1.1 \sigma$)  &   $  1 \pm   2$ ($ 0.6 \sigma$)  &   $  2 \pm   1$ ($ 1.5 \sigma$)  \\
 F110W  &   $ -1 \pm   2$ ($-0.8 \sigma$)  &   $  1 \pm   2$ ($ 0.3 \sigma$)  &   $  3 \pm   2$ ($ 1.7 \sigma$)  &   $  1 \pm   1$ ($ 0.8 \sigma$)  \\
 F125W  &   $ -0 \pm   2$ ($-0.3 \sigma$)  &   $  1 \pm   3$ ($ 0.4 \sigma$)  &   $  1 \pm   2$ ($ 0.4 \sigma$)  &   $  1 \pm   1$ ($ 0.4 \sigma$)  \\
 F140W  &   $ 10 \pm   2$ ($ 5.2 \sigma$)  &   $  9 \pm   2$ ($ 4.9 \sigma$)  &   $ 12 \pm   2$ ($ 5.5 \sigma$)  &   $ 11 \pm   1$ ($ 9.1 \sigma$)  \\
 F160W  &   $ 27 \pm   4$ ($ 7.4 \sigma$)  &   $ 25 \pm   4$ ($ 7.0 \sigma$)  &   $ 20 \pm   3$ ($ 7.4 \sigma$)  &   $ 24 \pm   2$ ($12.4 \sigma$)  \\
IRAC [3.6] &   62.0$\pm$  26.5 (2.3$\sigma$) &   14.4$\pm$  23.4 (0.6$\sigma$) &   57.5$\pm$  23.5 (2.4$\sigma$) &   44.6$\pm$  14.1 (3.2$\sigma$) \\
contamination\tablenotemark{2} & 166.9 (269\%) & 13.9 (97\%) & 27.6 (48\%) & \\
IRAC [4.5] &   62.6$\pm$  18.5 (3.4$\sigma$) &   15.5$\pm$  12.2 (1.3$\sigma$) &   53.2$\pm$  33.7 (1.6$\sigma$) &   43.8$\pm$  13.5 (3.2$\sigma$) \\
contamination\tablenotemark{2} & 135.8 (217\%) & 11.5 (74\%) & 21.8 (41\%) & 
\enddata
\tablenotetext{1}{All HST fluxes are adopted from \citet{coe13}, corrected for lensing magnification, and expressed in units of nJy. }
\tablenotetext{2}{The fluxes of neighbouring objects subtracted in the aperture divided by the magnification. These fluxes are also shown as fractions relative to the demagnified fluxes of JD1, JD2, and JD3 in parentheses.}
\end{deluxetable*}

\section{Photometry and Size Measurements}\label{sec:photometry}

\subsection{HST}\label{sec:hst}

Since the HST observations we make use of in the study of MACS0647-JD are identical to that in \citet{coe13}, for simplicity and consistency with the \citet{coe13} study, we make use of the same photometry presented in that study.
In \citet{coe13}, fluxes are measured in isophotal apertures using SExtractor \citep{bertin96}. 
The detection image is constructed from a weighted mean of the five WFC3 IR bands. 
The flux uncertainties are estimated based on the rms noise measured in apertures of similar size from the nearby background.  This approach directly accounts for the contribution of correlated noise in mosaic images.  Table \ref{table:photometry} lists the HST photometry after correction for the lensing magnification (see section \ref{sec:magnification}). 

\subsection{Spitzer/IRAC}\label{sec:irac}

The Spitzer/IRAC data we use in deriving fluxes for the various
multiple images of MACS0647-JD are based on individual exposures taken
with rotation angles such that the ``column pull-down'' artifact does
not overlap the sources on which we are performing photometry.  

Because of the high density of sources in the Spitzer/IRAC
observations, it is necessary to model and subtract flux from
neighboring sources before doing photometry on the faint multiple
images of interest.  As in previous work by our team \citep{labbe06,labbe10,labbe13,labbe15} and other teams
\citep{shapley05,grazian06,laidler07,merlin15}, we use the
high-resolution HST observations as a template for modeling the
lower-resolution Spitzer/IRAC observations.  
It has been shown by simulations that using a high-resolution prior greatly reduces the fraction of sources contaminated by neighbors in Spitzer/IRAC data.  \citet{labbe15} estimate the improvement in the fraction of contaminated sources to be as much as 80\% to 12\%.
In deriving a template for each source, we take directly the sextractor-segmented HST F160W
image, then convolve each segmented object with a mathematical mapping
from the HST PSF to the IRAC PSF. 

\begin{figure*}
\centering
\includegraphics[width=170mm]{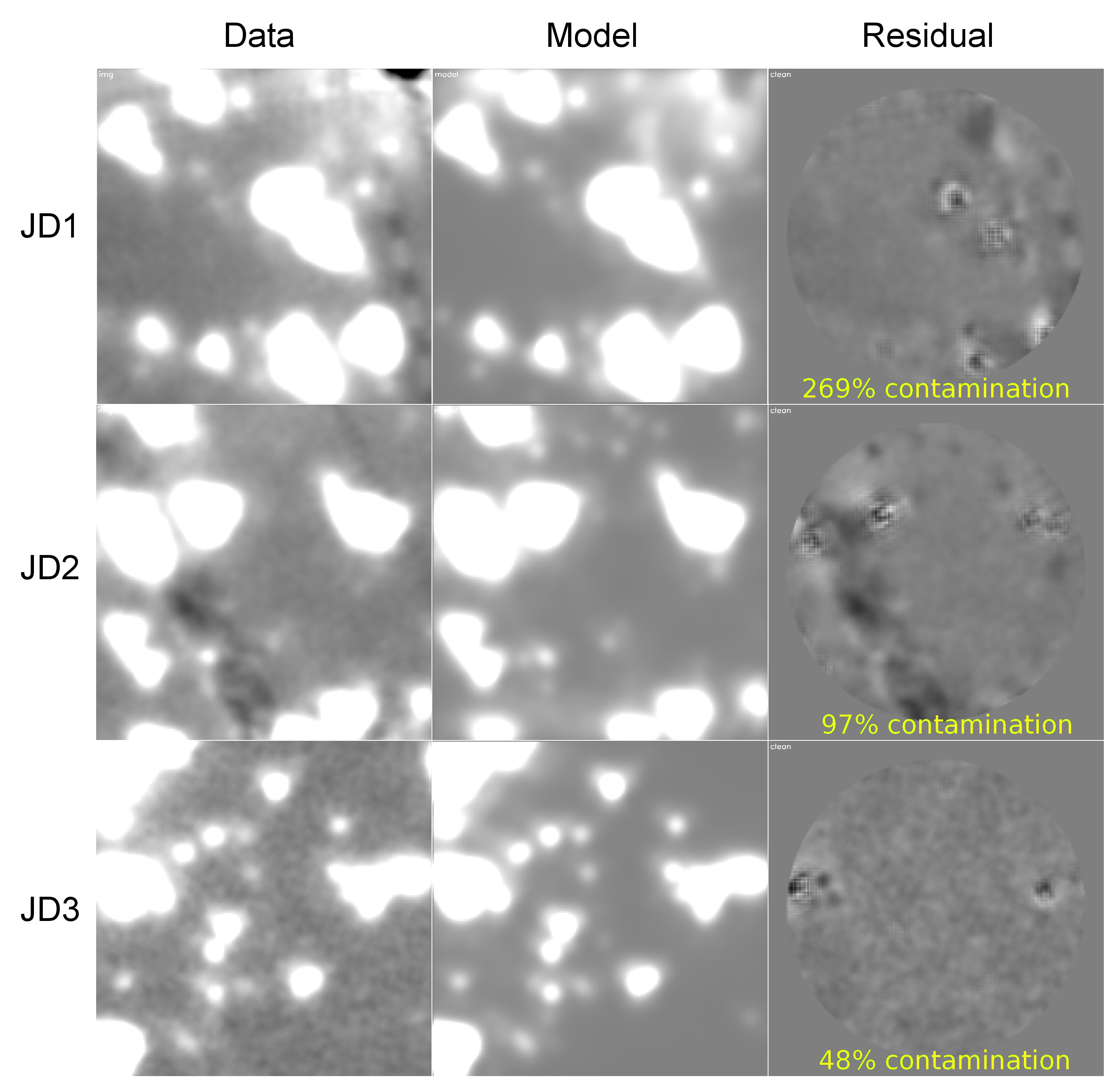}
\caption{Illustration of the view we obtain at $3.6\mu$m with
  Spitzer/IRAC of the various multiple images of MACS0647-JD, i.e.,
  JD1, JD2, and JD3 (\textit{shown in the upper, middle, and lower
    rows}), based on our model of the light from the neighboring
  sources.  The left column shows the original Spitzer/IRAC images,
  the middle column shows our model for the light from neighboring
  sources, and the right column shows the residual images after
  subtraction of all neighboring objects.  As the quality of the subtractions is
  similar in both the $3.6\mu$m and $4.5\mu$m, only the $3.6\mu$m
  image is shown here.  The relatively modest flux apparent in the
  Spitzer/IRAC imaging data, after subtraction of the light from the
  nearby neighbors, provides strong evidence that MACS0647-JD is
  unlikely to be a low-z interloper, consistent with the earlier
  conclusions of \cite{coe13} and \cite{pirzkal15}.}
\label{fig:mophongo_subtraction}
\end{figure*}

The IRAC PSF model is derived from a large number ($\approx$600) of warm
exposures, with spatial variation across the detector taken into
account.  We then select stars in the HST F160W image (26 of them in
this study), and compute the mathematical mapping to the IRAC PSF for
each star.  The mapping is parameterized by a number of basis
functions and their corresponding coefficients.  Using the selected
stars as anchors, the coefficients, are interpolated across the entire
field of view, which gives us a model of the PSF as a function of
spatial position.  Finally, the amplitude of each PSF mapped segment
is allowed to vary in order to minimize the residual.  Figure
\ref{fig:mophongo_subtraction} shows that most of the neighboring
sources around the targets are well-modeled and reasonably subtracted
from the data.  Thanks to both the depth of our new Spitzer/IRAC data and use of a more sophisticated photometry procedure, we are able to improve the quality of the Spitzer/IRAC photometry relative to that of \cite{coe13}.


We then measure the fluxes of our targets using circular apertures
with radius of 0.9" and then divide this flux by the expected fraction of
encircled energy inside this radius for a point source.  The fitting errors of
neighbouring objects are also taken into account, by adding in
quadrature with the Poisson noise measured by the apertures on the
source-subtracted image. 
The Spitzer/IRAC photometry are presented in Table \ref{table:photometry}. 

\begin{figure}
\centering
\includegraphics[width=85mm]{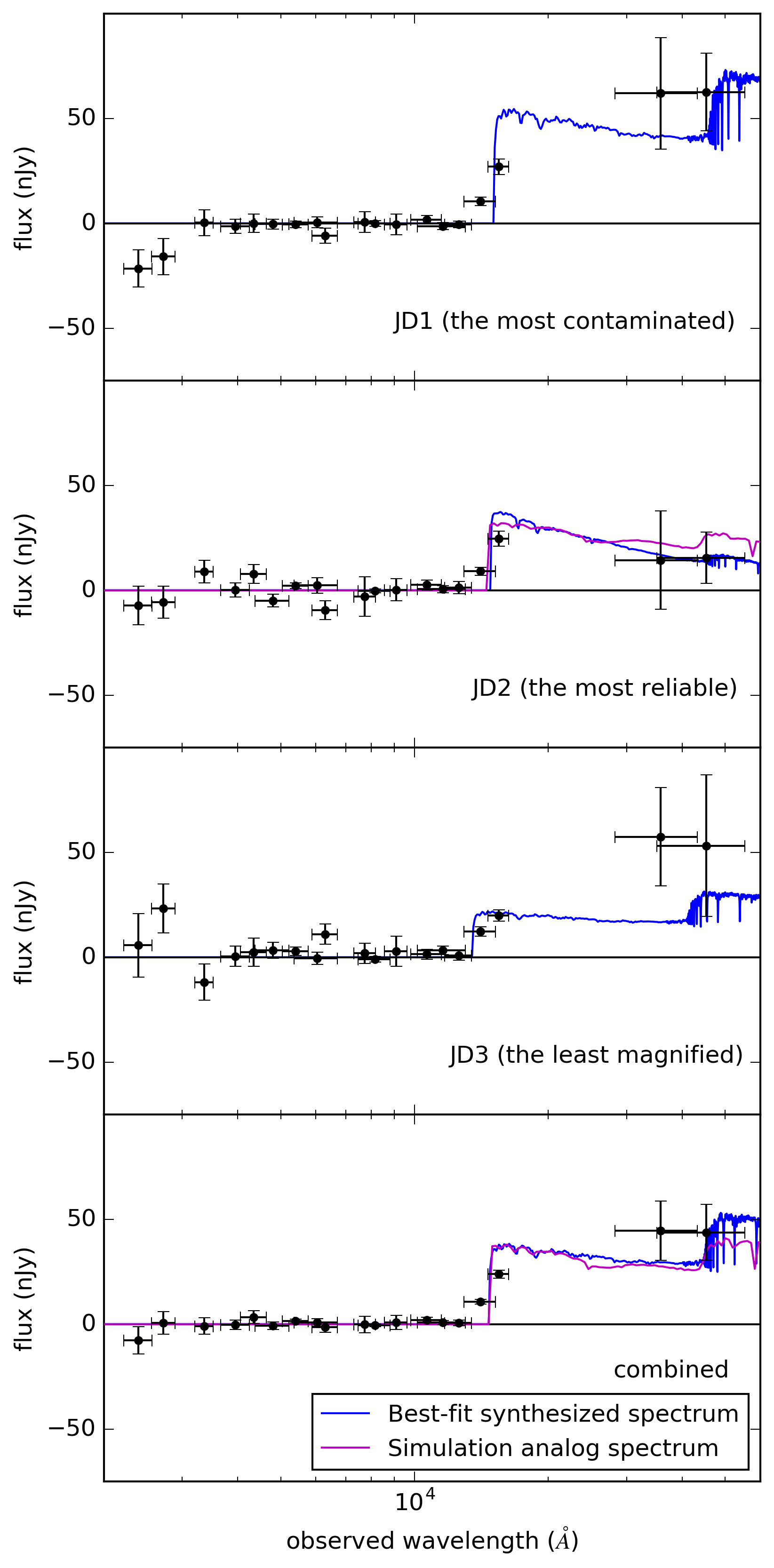}
\caption{ Best-fit stellar population model fits (\textit{blue lines})
  to the observed photometry (\textit{black points}) of the three
  multiple images, MACS0647-JD1, JD2, and JD3 (\textit{panels
    from the top}), as a function of the observed wavelength.  The
  lowest panel shows similar fits to the combined photometry for
  MACS0647-JD.  
  Shown for comparison are spectra from the EAGLE
  simulations that serve as analogues to EAGLE (\textit{magenta
    lines}).  These simulated spectra are slightly shifted from z=10 to z=11 and normalized to the best-fit synthesized spectra at rest-frame 1600 \r{A} to better match the observed spectra.\label{fig:sed_fit}}
\end{figure}

\subsection{Spectral Energy Distribution}

Figure \ref{fig:sed_fit}
plots the de-magnified spectral energy distribution (SED) of
MACS0647-JD1, JD2, JD3, and their stacked values.  Compared to JD1 and
JD3, our primary target, JD2, has a tentatively steeper rest-frame UV
slope, although still consistent with the measurements of the other
two multiple images.  Since most of the systematics have been removed,
this suggests the Poisson noise still dominate the uncertainty of
ultra-faint objects.

Of the three multiple images, JD2 is recognized as perhaps the best
target because of its high magnification and its being more separated
from bright neighbors.  JD1 is the most magnified but is also the most
contaminated (by a nearby cluster galaxy), while JD3 is the least
contaminated but also the faintest, having the lowest magnification factor
from gravitational lensing.

\subsection{Lensing Magnification}\label{sec:magnification}

The magnification factors of JD1, JD2, and JD3 are estimated by a suite of recently-constructed\footnote{These are improved versions of those available on the CLASH website: https://archive.stsci.edu/missions/hlsp/clash/macs0647/models/} light-traces-mass (LTM) models constructed using the methods of \citet{zitrin09} and \citet{zitrin14}. 
The models are constrained by multiply lensed images including those of MACS0647-JD. 
The uncertainties correspond to the 68.3\% confidence level from the model MCMC minimization, unless, these are exceeded by the magnifications in an area of 1"x1" around each multiple image, or the variation between different best-fit models probed in the modeling procedure. 

The magnification factors used in this work are different from those used by \citet{coe13}. 
Those are estimated from a Lenstool model, and are 8.4, 6.6, and 2.8 for JD1, JD2, and JD3, respectively. 
 

\subsection{Size Measurements}

We use the high spatial resolution HST data to measure physical sizes
for the various multiple images of MACS0647-JD.  We perform the
measurement using \textsc{galfit} \citep{peng10} and use it to fit
PSF-convolved S\'{e}rsic profiles to the F160W data.  The derivation
of the F160W PSF we utilize is described in \S\ref{sec:irac}.

Based on the fits we perform with \textsc{galfit}, we find that JD1
and JD2 are slightly resolved \citep[see also][]{coe13}.  Their measured
effective radii are 0.068$\pm$0.017" and 0.062$\pm$0.017", respectively, while we
find JD3 to be unresolved (i.e., $<$0.063$''$, 1-$\sigma$).  We converted the
observed image-plane sizes into physical sizes, by demagnifying each
source radius by the square root of the model-estimated magnification factor.
The results are presented in Table~\ref{table:properties}.
The mean intrinsic half-light radius measured for MACS0647-JD is 105$\pm$28 pc. 

Interestingly enough, the size we measure for MACS0647-JD is similar to the observed sizes of local star cluster complexes (e.g., 30 Doradus) and giant molecular clouds, which are on the order of 30-100 pc \citep{johansson98, murray11}.  The inferred stellar mass and star formation rate of MACS0647-JD (see \S\ref{sec:sps}) are much higher, of course, than local giant molecular clouds.  The fact that there are such large differences can be mainly attributed to the fact that mass density roughly scales as $(1+z)^{3}$ and thus star formation density was much, much higher in the early universe. 

\begin{deluxetable*}{ccrrrrrrc}
\tablecaption{\label{table:properties} Inferred properties of MACS0647-JD based on the measured photometry and light profile fitting for JD1, JD2, JD3, and 
a weighted average of the photometry for JD1, JD2, and JD3.}
\tablehead{
           \colhead{object} & 
           \colhead{z\tablenotemark{1}} &
           \colhead{M$_{UV,AB}$} &
           \colhead{log$_{10}$(age/yr)} & 
           \colhead{\pbox{20cm}{log$_{10}$(mass/\\M$_{\odot}$)}} & 
           \colhead{\pbox{20cm}{log$_{10}$(sfr/\\(M$_{\odot}$/yr))}} & 
           \colhead{\pbox{20cm}{rest-frame\\UV slope\\(measured\tablenotemark{2})}} & 
           \colhead{\pbox{20cm}{rest-frame\\UV slope\\(best-fit\\model\tablenotemark{1})}} &
           \colhead{\pbox{20cm}{intrinsic\\half-light radius\tablenotemark{3}\\(pc)}}
           }
\startdata
JD1  &  11.4$^{+0.4}_{-1.0}$  &  $-$20.3$\pm$0.3  &  8.6$^{+ 0.1}_{-2.2}$  &  9.2$^{+ 0.2}_{- 1.5}$  &   0.8$^{+0.9}_{-0.5}$  &  $-$1.0$\pm$0.8  &  $-$2.3  &  110$\pm$40 \\[5pt]
JD2  &  11.2$^{+0.4}_{-1.0}$  &  $-$19.6$\pm$0.3  &  6.8$^{+ 1.9}_{-1.8}$  &  7.7$^{+ 1.3}_{- 0.3}$  &  0.9$^{+1.1}_{-0.7}$  &  $-$2.7$\pm$1.7  &  $-$2.9 & 100$\pm$40 \\[5pt]
JD3  &  10.1$^{+0.9}_{-0.4}$  &  $-$20.4$\pm$0.5  &  8.7$^{+ 0.1}_{-3.7}$  &  8.8$^{+ 0.2}_{- 1.5}$  &  0.3$^{+1.2}_{-0.2}$  &  $-$0.7$\pm$0.9  &  $-$2.3 & $<$170 \\[5pt]
Stack  &  11.1$^{+0.5}_{-0.9}$  &  $-$20.1$\pm$0.2  &  8.6$^{+ 0.1}_{-2.1}$  &  9.1$^{+0.2}_{- 1.4}$  &  0.6$^{+0.7}_{-0.4}$  &  $-$1.3$\pm$0.6  &  $-$2.3
\enddata
\tablenotetext{1}{Estimated from the best-fit synthetic stellar population model assuming a constant star formation rate. See \S \ref{sec:sps} for a complete list of assumptions. }
\tablenotetext{2}{The rest-frame UV slope $\beta$, where $f_{\lambda} \propto \lambda^{\beta}$, measured directly from the HST F160W and Spitzer IRAC 3.6$\micron$ color \citep[as pioneered in][]{wilkins16}. }
\tablenotetext{3}{The intrinsic half-light radii are calculated assuming the source is at z=11 and corrected for lensing using the square root of the estimated magnification factors from the Zitrin-LTM model. }
\end{deluxetable*}

\section{Stellar population properties}\label{sec:sps}

\subsection{Stellar Population Modeling}

We derive the age, the mass, and the star formation rate, by fitting
synthesized stellar population spectra to the measured photometry
(Figure \ref{fig:sed_fit}) using FAST \citep{kriek09}.  For the
stellar population synthesis model, we choose \cite{bruzual03}.  This
particular model has a wider range in stellar ages (from 0.1 Myr to
the age of the Universe) than the other options \citep{maraston05,
  maraston11, conroy11}, which may be advantageous for modelling
galaxies at very high redshifts.  We adopt a Chabrier initial mass
function \citep{chabrier03}, which is derived from observations of
various parts of the Galaxy while the other two IMFs,
\cite{salpeter55} and \cite{kroupa02}, are derived from observations
of the solar neighborhood.

For simplicity, we assume a constant star formation history.  We adopt the dust law of \cite{kriek13}.  It
is a flexible attenuation law that varies with the spectral type of
galaxies, and is derived from galaxies at redshifts up to 2.  This
option is more flexible and is derived from galaxies at higher
redshifts compared to the other two options, \cite{calzetti00} and
\cite{cardelli89}.  For metallicity, we choose the lowest possible
value (Z = 0.004) given the expected young age of the galaxy at
z$\approx$11.  Finally, we keep the maximum possible extinction small,
A$_{v}$ = 0.1, as observed to be typical at such high redshifts for 
galaxies from the latest ALMA observations 
\citep{aravena16,bouwens16ir,dunlop16} of galaxies in the Hubble 
Ultra Deep Field \citep{beckwith06}.

\subsection{SED Fitting Results}

The physical properties we have inferred for each of the multiple
images of MACS0647-JD and the weighted average are compiled in table
\ref{table:properties}.  For its age, the three multiple
images-combined constraint is log$_{10}$(age/yr) = 8.6$^{+0.1}_{-2.1}$, while with the
optimal target - JD2 alone, the estimated age is substantially younger
- only log$_{10}$(age/yr) = 6.8$^{+1.9}_{-1.8}$, although still consistent within the
uncertainty.  For its stellar mass, the constraint based on the
combined photometry of the three multiple images is
log$_{10}$(M$_{*}$/M$_{\odot}$) = 9.1$^{+0.2}_{-1.4}$, while using only JD2,
the stellar mass is log$_{10}$(M$_{*}$/M$_{\odot}$) = 7.7$^{+1.3}_{-0.3}$.
The star formation rate based on the combined photometry is 
$4^{+16}_{-2}$ M$_{\odot}$ yr$^{-1}$ while the value based on the
JD2 photometry is $7^{+84}_{-6}$ M$_{\odot}$ yr$^{-1}$.  

As only the 4.5$\mu$m Spitzer/IRAC channel reaches the rest-frame optical and there are large uncertainties on the measured flux in that band, the age is not strongly constrained. 
This means that our stellar mass estimates are strongly correlated with the adopted stellar population age.  Specifically, if we consider only low values of the age, i.e., log$_{10}$(age/yr) $\approx$ 6.8, we infer stellar masses $\approx$1 dex lower than if higher values of the age, i.e., log$_{10}$(age/yr) $\approx$ 8.6, are considered. 

\begin{figure*}
\centering
\includegraphics[width=170mm]{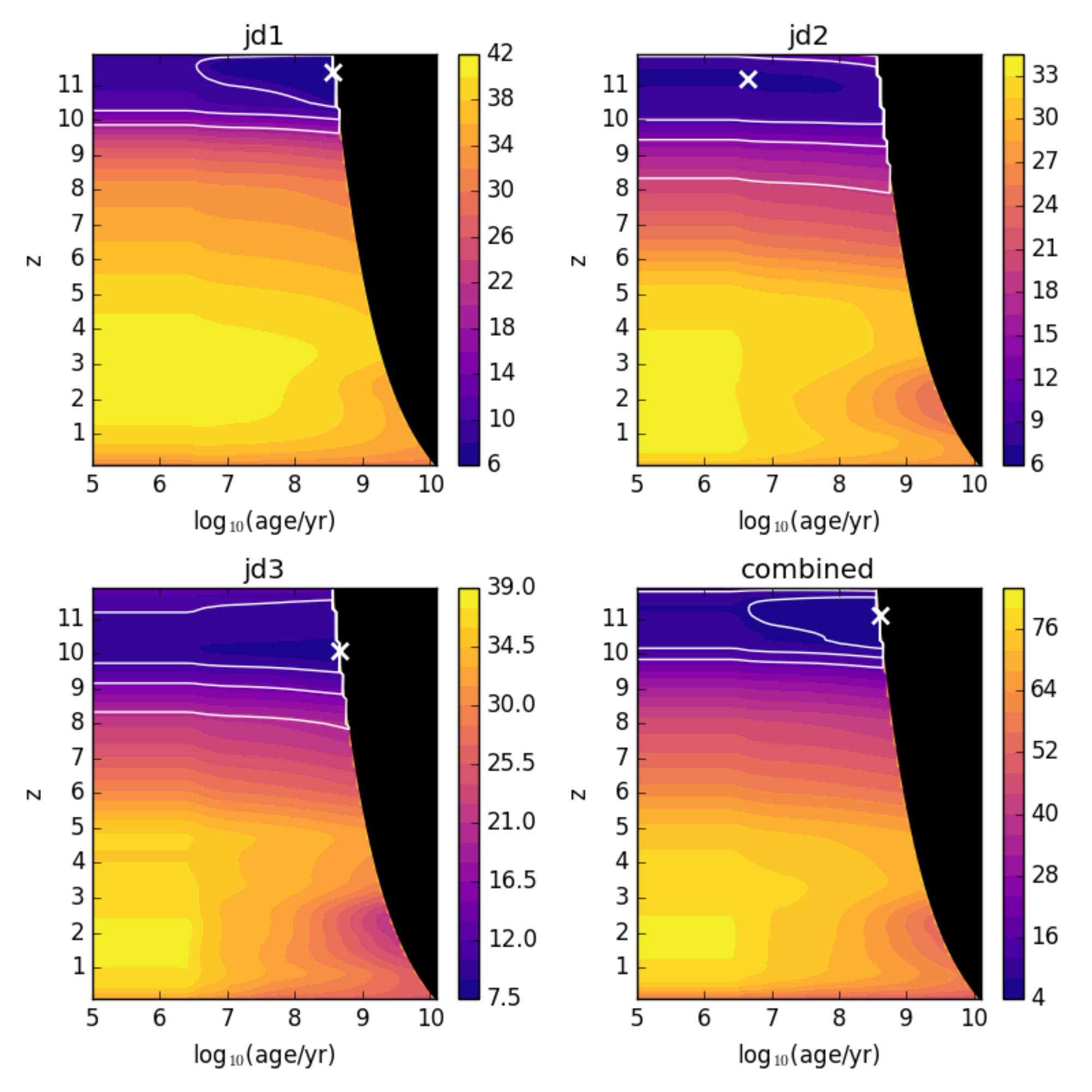}
\caption{Our $\chi^2$ fit results to MACS0647-JD vs. stellar
  population age and redshift, assuming a constant star formation rate.  Results are shown based on the
  photometry for JD1, JD2, JD3, as well as the combined photometry.
  The color bar provides the scale for the $\chi^2$ values.  The
  best-fit values of redshift and age are marked by the white
  crosses's. The white contours in each panel show the 1$\sigma$,
  2$\sigma$, and 3$\sigma$ confidence intervals.  The black region on the right side
  of each plot shows the disallowed region of parameter space, where
  the stellar population ages exceed the age of the Universe at the
  corresponding redshifts.\label{fig:chi_plot}}
\end{figure*}

The new constraints we have obtained from SED-fitting are similar to the first-order estimates of \citet{coe13}, who derived a star formation rate of $\approx$ 4 $M_{\odot}$ yr$^{-1}$ and a stellar mass on the order of $10^{8}-10^{9} M_{\odot}$. 
Figure \ref{fig:chi_plot} shows the $\chi^2$ fit values for JD1, JD2, JD3
individually and the combined constraint. 

The absolute UV magnitudes, $M_{UV}$, of JD1, JD2, and JD3 are $-$20.49, $-$19.91, and $-$19.32, respectively. 
Their average value is $-$19.91. 
We assume the best-fit redshifts from FAST, which are z=11.4, z=11.2, and z=10.1 for JD1, JD2, and JD3, respectively. 
$M_{UV}$ is measured at a rest-frame wavelength of 1600\r{A}. 
Since the observed wavelength ($\approx$1.9 $\mu$m) is not covered by any filters, we infer $M_{UV}$'s from FAST's best-fit spectra. 

Using the H$_{160}-$[3.6] color \citep{wilkins16}, we find that the rest-frame UV slope,
$\beta$, of the stacked photometry is $-$1.3$\pm$0.6, and is
$-$2.7$\pm$1.7 for JD2 alone.  Calculating $\beta$
directly from the color of F160W and IRAC [3.6], however, does not
take into account of the Lyman break shifting into the F160W filter
which occurs at z$\approx$10.5.  Another way to estimate $\beta$ is to
take the values from the best-fit stellar population models, which are
found to be $-$2.3 and $-$2.9 for the stacked photometry and JD2
respectively.  These estimates for $\beta$ are mostly bluer than our
estimates based on colors themselves, but are highly consistent with
the age and dust extinction we estimate for the source.

The use of different stellar population parameters to model the
photometry of MACS0647-JD results has only a modest impact on the
result.  Use of another stellar-population model
\citep[e.g.][]{conroy11}, dust model \citep[e.g.][]{calzetti00}, or
metallicities (e.g. solar metallicity: Z = 0.02) only results in
$\approx$10 - 20$\%$ changes in the derived properties.  If we do allow a
larger maximum possible extinction of A$_{v}$ = 1.0, then the age can
decrease by $\approx70\%$ while the mass increases by $\approx30\%$, which
together can cause the estimated star formation rate to increase by a
factor of 3.  As expected, use of a \cite{salpeter55} IMF
significantly increases the stellar mass ($\approx70\%$) we infer for
MACS0647-JD.

The stellar population modeling we have performed here should benefit
from significantly improved constraints on the redshift of MACS0647-JD
from HST grism data.  These data included 12 orbits of HST grism data
obtained as part of a cycle 21 program (GO-13317, PI: Coe).
Fixing the redshift to z=11.2 and z=11.1 for JD2 and the combined photometry (their best-fit redshift derived by FAST), we arrive at tighter constraints on the
derived age.  Using the JD2 and combined photometry, we derive an age
of log$_{10}$(age/yr) = 6.8$^{+1.0}_{-1.8}$ and log$_{10}$(age/yr) = 8.6 (errors less than 0.1 dex when redshift is fixed to $z=11.1$), respectively.

The \citet{bruzual03} models used by FAST does not include nebular emission line models. 
\citet{Pacifici2015} showed that emission lines contamination have a negligible impact on stellar mass estimates for  $z\approx1-3$ star-forming galaxies.  However, at $z\approx10$, the star-formation/ISM conditions of galaxies could be extreme, resulting in a significant contribution to the photometry by strong  Lyman-$\alpha$ and other rest-UV ISM emission lines.  
This would result in the FAST derived galaxy properties being systematically biased. 
In order to constrain the impact of nebular emission lines on the derived galaxy properties, we use the Flexible Stellar Population Synthesis code \citep{conroy2009} within the Prospector framework \citep{leja2017} to obtain best-fit galaxy properties via SED fitting. 
We use an exponentially declining SFH with an e-folding time of 1 Gyr with redshift fixed at $z=11.1$ and vary the stellar mass, stellar metallicity, age, and dust extinction (following \citealt{calzetti00}) between reasonable parameter space with a tophat prior. 
The median deviation for the age and stellar metallicity estimates between models with and without nebular contribution (emission + continuum) is consistent with no statistically significant difference. 
However, the tests we run suggest the stellar mass estimates would be systematically overestimated by $\approx0.7\pm0.3$ dex when emission lines are not included. 
Additionally, there is also a slight tendency for the dust extinction to be overestimated when not accounting for the strong nebular lines.  We note that our observational data is limited to $\lambda<4.5\mu m $, thus we have no constraints on the rest-frame optical and NIR spectral shape of MACS0647-JD, which is crucial for reliable stellar mass/metallicity estimates.

\begin{figure}
\centering
\includegraphics[width=85mm]{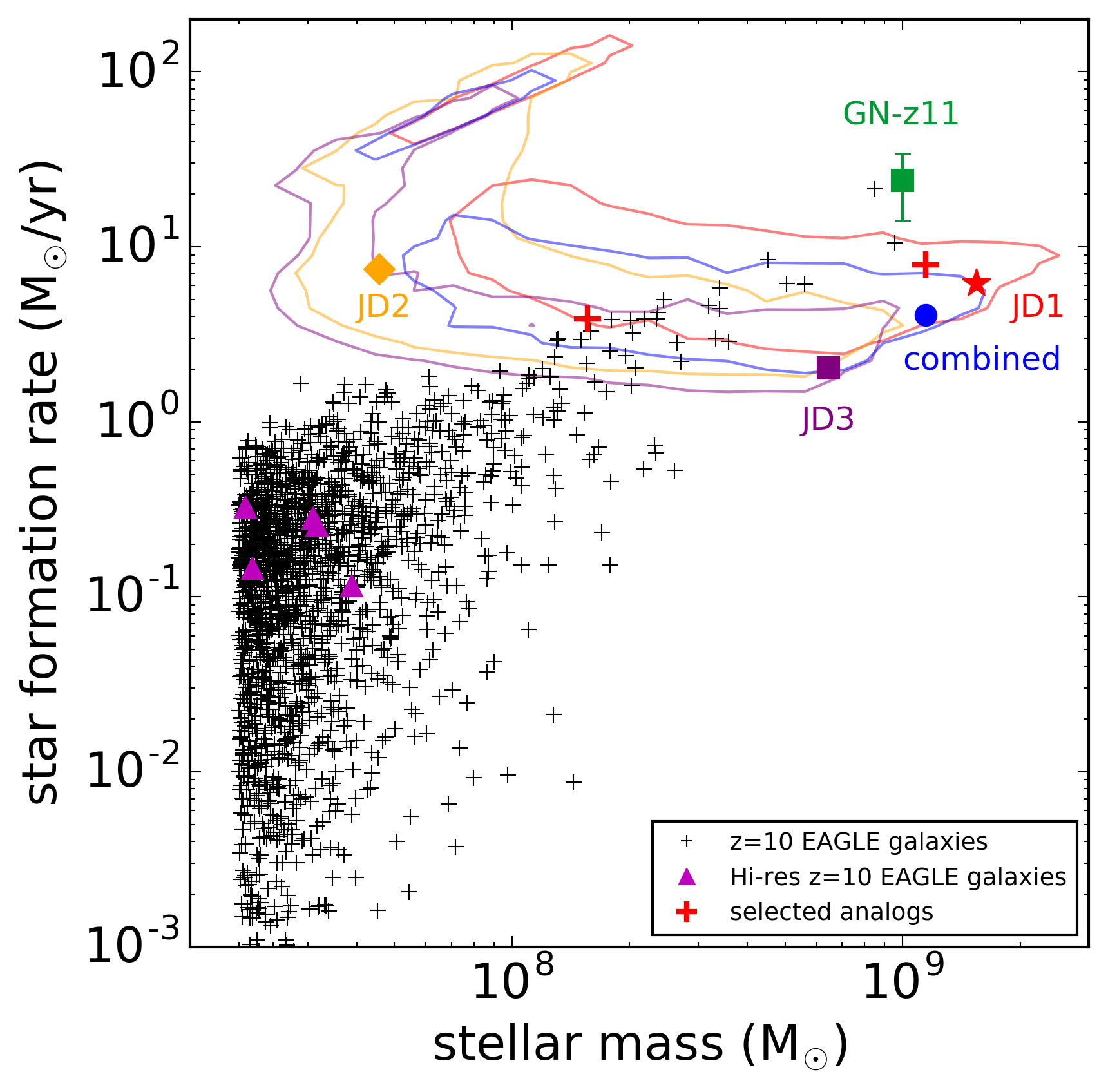}
\caption{SFR vs. stellar mass inferred for various multiple images of
  MACS0647-JD based on stellar-population modeling of the observed
  photometry.  Separate results are plotted for each of the multiple
  images, i.e., JD1 (\textit{red star}), JD2 (\textit{orange
    diamond}), JD3 (\textit{purple square}), and a weighted average of
  the photometry for all three multiple images (\textit{blue cross}).
  Their 1-$\sigma$ confidence levels are plotted as solid lines with the same colors. 
  Also plotted for context are the recent $z=11.1$ galaxy GN-z11 reported by
  \cite[green square]{oesch16} and $z\approx10$ galaxies (\textit{black crosses}) from
  EAGLE (corresponding to the closest snapshot in EAGLE to
  $z\approx11.1$).  Galaxies at the same redshift found in a smaller
  volume but higher resolution version of the EAGLE simulation are
  denoted by magenta triangles.  The red crosses denote the galaxies
  from EAGLE whose properties most resemble those of JD2 and the
  weighted average of each of the multiple images.  The inferred properties for MACS0647-JD seem consistent with some of the  most massive and the most rapidly star-forming galaxies in the EAGLE simulation.\label{fig:sfr_vs_mass}}
\end{figure}

\begin{figure}
\centering
\includegraphics[width=85mm]{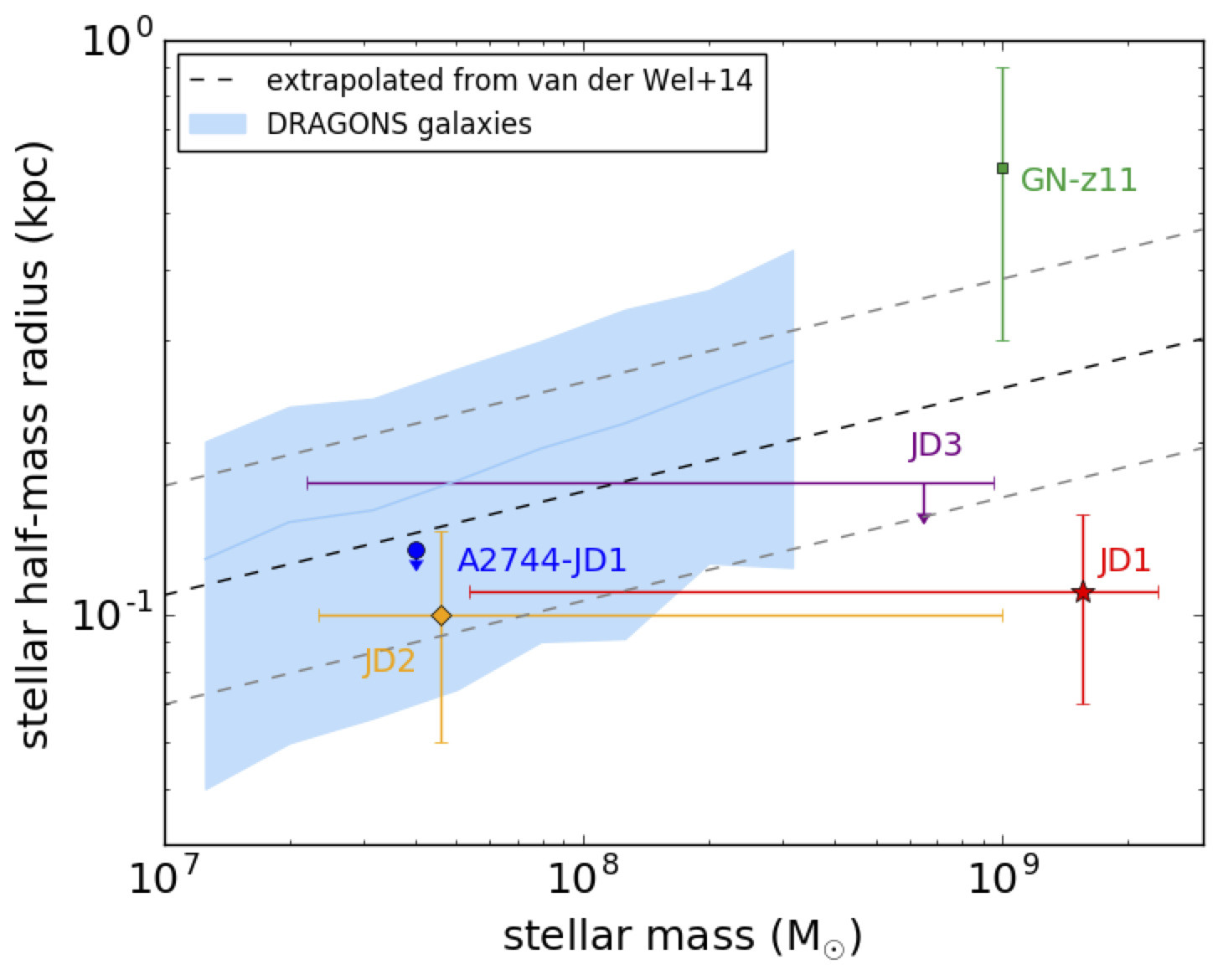}
\caption{Stellar mass vs. half-mass radius derived for the
  different multiple images of MACS0647-JD, the $z=11.1$ galaxy
  GN-z11 (green square), and the z$\approx$10 galaxy A2744-JD1 \citep[][blue circle]{zitrin14} compared against similar quantities in the semi-analytical DRAGONS 
  simulation.  The light blue line and the shaded area denote 
  the average sizes and the 1-$\sigma$ dispersion of DRAGONS galaxies
  as a function of stellar mass at z$\approx$10.9.  The size-mass relation of DRAGONS galaxies
  is shown up to $\approx$5$\times$10$^{8}$M$_{\odot}$, beyond which the number of galaxies
  becomes very small. 
  Overplotted as the black dashed line is
  the size-mass relation of late-type galaxies at $z=2.75$ by
  \cite{vanderwel14}, extrapolated to $z=11$ using the best-fit
  scaling with redshift z=2.75 also found in \cite{vanderwel14} study. 
  The gray dashed lines denote the scatter in the relation extrapolated to $z=11$. 
\label{fig:size_mass}}
\end{figure}

\begin{figure*}
\centering
\includegraphics[width=130mm]{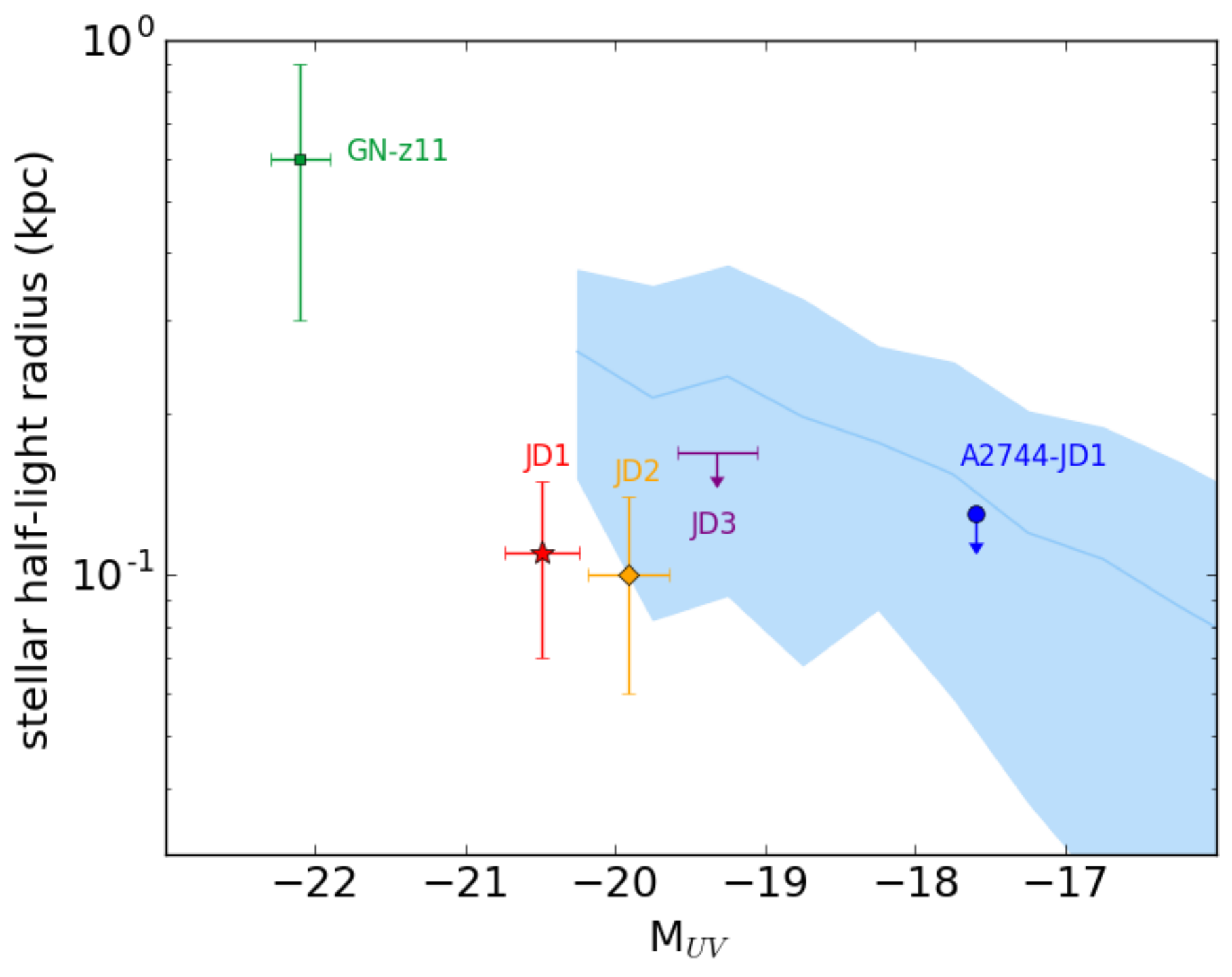}
\caption{Half-light radius vs. absolute UV magnitude inferred for each multiple image of MACS0647-JD, GN-z11 (\textit{green square}), and A2744-JD1 (blue circle).  These current observational results for z $\approx$ 10 – 11 galaxies are shown relative to the distribution in parameter space seen in the DRAGONS \citep{liu17} simulations, with the blue shaded region representing the $\pm1\sigma$ sizes in the simulations and the blue line representing the average size-M$_{UV}$ relation.  The smaller than expected sizes of low luminosity sources at $z\approx11$ recalls the very small size results at $z\approx6$-9 \citep{bouwens18,kawamata15, kawamata18}, where many sources with $\leq$100 pc sizes are found.\label{fig:size_muv}}
\end{figure*}

\section{Comparison with cosmological simulation}\label{sec:eagle}

It is interesting to compare the present observational results on
MACS0647-JD with that found in current state-of-the-art cosmological
hydrodynamical simulations.  One such simulation is the EAGLE project
\citep{schaye15}.  The simulation successfully captures a diverse
range of astrophysical phenomena using various subgrid physics
recipes, including radiative cooling, star formation, stellar
feedback, AGN feedback, accretion and mergers of supermassive black
holes.  
The amplitude of these effects are tuned to reproduce the observed galaxy stellar mass function, the relation between stellar mass and halo mass, galaxy sizes, and the relation between black hole mass and stellar mass at $z\approx0$.  
Encouragingly, the tuned simulation is found to reproduce a number of low-redshift observations not used in setting the subgrid physical prescriptions.  These observations include the specific star formation rates and fractions of passive galaxies, the relation between the maximum rotation speed and stellar mass (Tully-Fisher relation), the relations between metallicity (of ISM and stars) and stellar mass, the relation between the luminosity and mass of galaxy clusters, the relation between X-ray luminosity and temperature of the intracluster medium, as well as the column density distributions of intergalactic metals.  One might therefore ask whether the physical properties we infer for our $z\approx11$ candidate are a good match to galaxies in the EAGLE simulation. 

We select the analogs from the main run of EAGLE, which simulates a
volume of 100$^{3}$ comoving-Mpc$^{3}$ with a gravitational resolution
of 2.66 kpc (comoving).  The EAGLE analogs are selected using the criteria that
they have properties (SFR and stellar mass) that are most consistent
with that inferred based on the JD2 photometry (the ideal target) as
well as on the combined photometry.  Figure \ref{fig:sfr_vs_mass} plots
the star formation rate against the stellar mass for MACS0647-JD and
EAGLE galaxies taken from the z = 10 snapshot with stellar masses
greater than 2$\times$10$^{7}$ M$_{\odot}$.  The EAGLE simulation is
captured in 29 snapshots from z = 20 to z = 0.  The snapshot at z = 10
is the most relevant one to this work.  We also include galaxies found in
a variant of the EAGLE main run that has a smaller volume but higher
resolution into the analog selection.  The galaxies found in the
smaller simulation volume, however, tend to be the less massive and
less rapidly star-forming.

Radiative transfer is not implemented within the EAGLE simulation.
Rendering images and spectra of EAGLE galaxies requires
post-processing of the particle information.  In this case we use
SKIRT \citep{camps15}, a radiative transfer with dust processing code
that takes snapshots from hydrodynamical simulations as input, to
simulate the images and spectra of the EAGLE analogs.  Specifically,
the position, stellar mass, and age of star particles of the selected
analogous galaxies from EAGLE are used as inputs.  We compute results
using \cite{bruzual03} stellar population synthesis libraries and the
\cite{draine07} dust model.

The spatially integrated spectra are overplotted in magenta in Figure \ref{fig:sed_fit} on JD2 and the stacked photometry.  
The spectra are slightly shifted from z=10 to z=11 and normalized to the best-fit synthesized FAST spectra at rest-frame 1600 \r{A} to better compare their shapes. 
The spectral shapes of the analogs from the EAGLE simulation resemble those seen in the various multiple images of MACS0647-JD.

In terms of stellar population properties, the selected analogs of
MACS0647-JD are relatively high mass and with a high SFR relative to
the general population found in the EAGLE simulation.  This is not
surprising given that the observation strategy of CLASH to survey a
large number of galaxy cluster lenses at modest depths.

It would not be very meaningful to compare the observed sizes of MACS0647-JD
with sources in the EAGLE simulation, since EAGLE lacks the necessary
spatial resolution.  At z = 11, the gravitational softening length of
EAGLE is $\approx$0.2 proper kpc \citep{schaye15}, which is greater than
the measured sizes of the multiple images JD1 and JD2 (Table
\ref{table:properties}).  We therefore compare our measured sizes with
predictions made by another cosmological simulation, DRAGONS
\citep{liu17}. 
DRAGONS is a semi-analytical galaxy formation simulation designed to study the galaxy formation and the structure of reionization in the early universe at redshifts between z=35 and z=5 \citep{mutch16}.  While EAGLE simulates dark matter and gas particles in a fully self-consistent way, the galaxy formation and ionization parts of DRAGONS \citep{mutch16} are semi-analytical models based on a collisionless N-body cosmological simulation called \textit{Tiamat} \citep{poole16}.  
These physical models are calibrated to a hydrodynamical cosmological simulation called \textit{Smaug} \citep{duffy14}, which simulates a much smaller volume (10 $h^{-1}$ cMpc each side, but includes important physics like the interaction between radiation and gas. 
The overall less computational-intensive scheme of DRAGONS means it can simulate features that EAGLE lacks, such as the interaction between radiation and baryons.   It also means that DRAGONS has a higher spatial resolution.  At z=11, the gravitational softening length is $\approx$75 proper pc \citep{poole16}, which is slightly smaller than the radius we infer for MACS0647-JD. 

In Figure \ref{fig:size_mass} we plot the stellar half-mass radius
against the stellar mass for MACS0647-JD and DRAGONS galaxies.  We also consider an extrapolation of the size-mass relation derived for $0 < z < 3$
late-type galaxies from the 3D-HST and CANDELS program
\citep{vanderwel14}.  This size-mass relation is extrapolated from
z=2.75 to z=11 using the size-mass redshift evolution derived in that
study.  Specifically, it is 
\begin{equation}
R_{\textrm{eff}}/\textrm{kpc} = 10^{\textrm{A}} (M_{*}/5\times10^{10} M_{\odot})^{\alpha} ((1+z)/3.75)^{-0.75} \textrm{,} 
\end{equation}
where A = 0.51$\pm$0.01, $\alpha$ = 0.18$\pm$0.02, and z = 11.  
Note that the power-law dependence on $(1+z)$ for the size-mass relation is not especially surprising given that $\rho \propto (1+z)^{3}$ and hence $r \propto M^{1/3} (1+z)^{-1}$ \citep[e.g.,][]{ferguson04}. 
The observed size of the different multiple images of MACS0647-JD is generally smaller than expected, but the comparison is challenging due to significant uncertainties in our stellar mass estimates.

A more direct evaluation of the observed sizes of MACS0647-JD relative to that seen in the simulations can be achieved by conducting the comparisons at a fixed UV luminosity (Figure~\ref{fig:size_muv}).  It is clear that MACS0647-JD is $\approx$2$\times$ smaller in terms of its size at $M_{UV,AB}\approx-20$ mag than sources in the DRAGONS simulation.  The present results bring to mind the unexpectedly small observed sizes found by \cite{bouwens17,bouwens18} and \cite{kawamata15,kawamata18} from the Hubble Frontier Fields \citep{lotz17} or the proto-globular cluster candidates from \cite{vanzella17}.  In the former studies \citep[see also][]{ma18}, it was suggested that the small sizes likely resulted from just a fraction of a galaxy lighting up with star formation.  However, with only one size measurement for a $>$$-$21 mag galaxy at $z\approx11$, we are clearly very limited in the conclusions we can draw.


\section{Summary}

We describe the use of new, deep Spitzer/IRAC observations
we obtained of a particularly bright $z\approx$11.1 galaxy MACS0647-JD to
constrain its physical properties.  MACS0647-JD benefits from
substantial lensing magnification from the galaxy cluster MACS0647,
with the three multiple images with estimated magnification factors of
6.0, 5.5, and 2.1.  The source is unique in that it is one of two
moderately bright galaxy candidate known at $z\approx$11 -- the other
being GN-z11 \citep{oesch16} -- which are amenable to such detailed
study.

The new Spitzer/IRAC observations (50 hours per band: PI Coe) probe
much fainter than the previously existing 5-hour observations of this
candidate.  The deeper observations allow us to place improved
constraints on its stellar mass, age, and dust properties.  After
subtracting the flux from neighboring sources and combining the three
different multiple images, we obtain a secure $>$3$\sigma$ detection
of the source in the Spitzer/IRAC data.

We then derive constraints on its stellar population properties,
including age, stellar mass, star formation rate, and rest-frame UV
slope.  The age, stellar mass, and rest-frame UV slope we estimate for
MACS0647-JD based on a stack of the photometry are log$_{10}$(age/yr) = 8.6$^{+0.1}_{-2.1}$, log$_{10}$(M$_{*}$/M$_{\odot}$) = 9.1$^{+0.2}_{-1.4}$, and $-$1.3$\pm$0.6,
respectively.  Using the least-contaminated and most magnified image
yields lower but still consistent values of log$_{10}$(age/yr) = 6.8$^{+1.9}_{-1.8}$,
log$_{10}$(M$_{*}$/M$_{\odot}$) = 7.7$^{+1.3}_{-0.3}$, and $-$2.7$\pm$1.7.

We compared the stellar mass and star formation rate we derive for MACS0647-JD with predictions for the source from the cosmological hydrodynamical simulation, EAGLE, and find that MACS0647-JD resembles the more massive and rapidly star-forming ones in EAGLE. 

We also measure the physical half-light radius of MACS0647-JD and find 105$\pm$28 pc.  Interestingly enough, the observed size we measure is $\approx 2 \times$ smaller than the mean sizes of $\approx-$20 mag sources in the DRAGONS simulations.  The very small observed sizes of MACS0647-JD (vs. expectations) is evocative of similarly small sizes of $z\approx$6-10 sources recently reported behind the HFF clusters \citep{kawamata15,bouwens17,bouwens18,kawamata18,vanzella17,zitrin14}.  Comparisons with observed sizes in EAGLE are not useful, given that the larger spatial extent of the softening length relative to the observed sizes at $z\approx 11$.

It is remarkable that we are able to map out the overall shape of the SED of a galaxy (to $\approx$0.4$\mu$m) at $z\approx11$ with HST and Spitzer Space telescopes and to constrain its stellar population age, stellar mass, and dust content.  Unfortunately, we are limited in terms of the conclusions we could draw from this one source, and so we will be conducting follow-up analyses on a larger sample of lensed sources in the epoch
of reionization (z $>$ 6) in datasets such as CLASH \citep{postman12}, Hubble Frontier Fields \citep{lotz17}, and SURFS UP \citep{bradac14}.

\acknowledgments 
We gratefully acknowledge funding from NOVA.   This work is based [in part] on observations made with the Spitzer Space Telescope, which is operated by the Jet Propulsion Laboratory, California Institute of Technology under a contract with NASA. Support for this work was provided by NASA through an award issued by JPL/Caltech.
DL thanks the contribution from Chuanwu Liu and Stuart Wyithe on the DRAGONS simulation.

\clearpage

\end{document}